\documentclass[aps,prb,superscriptaddress,twocolumn]{revtex4}
\usepackage{amssymb}
\usepackage{graphicx}

\begin{document}
\title{Nature of the Magnetic Order in BaMn$_2$As$_2$}
\author{Yogesh Singh}
\affiliation{Ames Laboratory and Department of Physics and Astronomy, Iowa State University, Ames, Iowa 50011,USA}
\author{M. A. Green}
\affiliation{NIST Center for Neutron Research, National Institute of Standards and Technology, Gaithersburg, Maryland 20899, USA}
\affiliation{Department of Materials Science and Engineering, University of Maryland, College Park, Maryland 20742, USA }
\author{Q. Huang}
\affiliation{NIST Center for Neutron Research, National Institute of Standards and Technology, Gaithersburg, Maryland 20899, USA}
\affiliation{Department of Materials Science and Engineering, University of Maryland, College Park, Maryland 20742, USA }
\author{A. Kreyssig}
\affiliation{Ames Laboratory and Department of Physics and Astronomy, Iowa State University, Ames, Iowa 50011,USA}
\author{R. J. McQueeney}
\affiliation{Ames Laboratory and Department of Physics and Astronomy, Iowa State University, Ames, Iowa 50011,USA}
\author{D. C. Johnston}
\affiliation{Ames Laboratory and Department of Physics and Astronomy, Iowa State University, Ames, Iowa 50011,USA}
\author{A. I. Goldman}
\affiliation{Ames Laboratory and Department of Physics and Astronomy, Iowa State University, Ames, Iowa 50011,USA}
\date{\today}

\begin{abstract}
Neutron diffraction measurements have been performed on a powder sample of BaMn$_2$As$_2$ over the temperature $T$ range from 10~K to 675~K\@. These measurements demonstrate that this compound exhibits collinear antiferromagnetic ordering below the N\'eel temperature $T_{\rm N}= 625(1)$~K\@.  The ordered moment $\mu = 3.88(4)~ \mu_{\rm B}$/Mn at $T$~=~10~K is oriented along the $c$ axis and the magnetic structure is $G$-type, with all nearest-neighbor Mn moments antiferromagnetically aligned.  The value of the ordered moment indicates that the oxidation state of Mn is Mn$^{2+}$ with a high spin $S = 5/2$.  The $T$ dependence of $\mu$ suggests that the magnetic transition is second-order in nature.  In contrast to the closely related $A$Fe$_2$As$_2$ ($A =$~Ca, Sr, Ba, Eu) compounds, no structural distortion is observed in the magnetically ordered state of BaMn$_2$As$_2$.  Our results indicate that while next-nearest-neighbor interactions are important in the $A$Fe$_2$As$_2$ materials, nearest-neighbor interactions are dominant in BaMn$_2$As$_2$. 

\end{abstract}

\maketitle

\begin{figure}[t]
\includegraphics[width=1.25in]{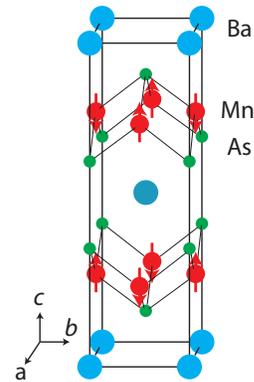}
\caption{(Color online) The crystallographic and magnetic structures of BaMn$_2$As$_2$.  The arrows on the Mn atoms represent the $G$-type arrangement of the Mn$^{2+}$ spins in the antiferromagnetically ordered state. 
\label{Fig-structure}}
\end{figure}

BaMn$_2$As$_2$ crystallizes in the tetragonal ThCr$_2$Si$_2$-type structure, shown in Fig.~\ref{Fig-structure}, and is isostructural to the $A$Fe$_2$As$_2$ ($A$~=~Ba, Sr, Ca, and Eu) family of compounds which display coupled antiferromagnetic (AF) and structural transitions.\cite{Rotter2008,Krellner2008,Ni2008,Yan2008,Ni2008a,Ronning2008,Goldman2008,Tagel2008,Ren32008,Jeevan2008}  Superconductivity is observed in the $A$Fe$_2$As$_2$ compounds on doping at the $A$ site,\cite{2Rotter2008,2GFChen2008,2Jeevan2008,Sasmal2008} by in-plane doping at the Fe site,\cite{Sefat2008, Jasper2008,Li2008} or by application of external pressure.\cite{Igawa2009, Kotegawa2009, Mani2009, Alireza2009}
The magnetic, thermal, and electronic properties of single crystals of BaMn$_2$As$_2$ have been reported recently.\cite{An2009,Singh2009}  In contrast to the $A$Fe$_2$As$_2$ materials, BaMn$_2$As$_2$ has an insulating ground state.\cite{An2009,Singh2009}  Magnetization measurements on single crystals suggested that BaMn$_2$As$_2$ has a collinear AF ground state with the easy axis along the $c$ axis and with a high N\'eel temperature $T_{\rm N} > 395$~K, and shows no structural distortion at 300~K\@.\cite{Singh2009}  BaMn$_2$As$_2$ is, therefore, unique compared to the other Ba$M_2$As$_2$ ($M$~=~Cr, Fe--Cu) compounds which are all metals with itinerant magnetic interactions or ordering.\cite{Yildirim2009}  We have previously suggested\cite{Singh2009} that the magnetic and electronic properties of BaMn$_2$As$_2$ are intermediate between those of the itinerant antiferromagnets $A$Fe$_2$As$_2$ mentioned above,\cite{Yildirim2009} and the local moment antiferromagnetic insulator La$_2$CuO$_4$,\cite{Johnston1997} both of which are parent compounds for high temperature superconductors.  
It is thus important to determine the actual magnetic structure of BaMn$_2$As$_2$, the value of the ordered moment, the N\'eel temperature, and the thermodynamic order (continuous or discontinuous) of the magnetic phase transition, and compare these properties with the magnetism found in the $A$Fe$_2$As$_2$ and La$_2$CuO$_4$ compounds to try to understand the relation between magnetism and superconductivity in these materials.

Herein we report neutron diffraction measurements on a powder sample of BaMn$_2$As$_2$ that answer these questions.  We find that BaMn$_2$As$_2$ becomes antiferromagnetically ordered below a high N\'eel temperature $T_{\rm N}= 625(1)$~K and with a refined ordered moment of $3.88(4)~ \mu_{\rm B}$ (at temperature $T$~=~10~K) oriented along the $c$ axis.  The magnetic structure is found to be $G$-type, a collinear antiferromagnetic structure in which nearest-neighbor spins in the tetragonal basal plane are antiparallel and successive planes along the $c$ axis are also antiferromagnetically aligned.  The temperature dependence of the ordered moment suggests that the magnetic ordering is second-order in nature.  There is no detectable structural transformation or distortion in the magnetically ordered state.  These properties will be discussed in terms of those of the $A$Fe$_2$As$_2$ compounds.
     
A polycrystalline sample (4.8~g) of BaMn$_2$As$_2$ was prepared through solid state synthesis by reacting small pieces of Ba metal with prereacted MnAs taken in the ratio Ba:MnAs~=~1.05~:~2.   Extra Ba was used in the starting composition to compensate for the loss of Ba due to evaporation and also to avoid the formation of the MnAs phase (a ferromagnet with a Curie temperature $T_{\rm C} \approx$~320~K).\cite{Bean1962}  The mixture was pelletized, placed in an alumina crucible and sealed in an evacuated quartz tube.  The tube was placed in a box furnace and heated to 850~$^\circ$C at a rate of 100~$^\circ$C/h, maintained at this temperature for 18~h, furnace cooled to 500~$^\circ$C,  and then air quenched.  The tube was opened in an argon-filled glove box and the resultant material was reground, pelletized and sealed in a quartz tube under vacuum again.  The tube was placed in a box furnace and heated to 750~$^\circ$C at a rate of 100~$^\circ$C/h, allowed to react for 6~h, heated to 900~$^\circ$C at a rate of 100~$^\circ$C/h, maintained at this temperature for 16~h, furnace cooled to 500~$^\circ$C\@ and then air quenched from this temperature.  This process was repeated twice.  Powder x-ray diffraction of the final product showed 92 wt.\% of the BaMn$_2$As$_2$ phase and about 8 wt.\% of the impurity phase, Ba$_2$Mn$_3$As$_2$O$_2$, a compound whose room temperature crystal structure and magnetic properties are known.\cite{Brechtel1979,Brock1996} 

The powder neutron diffraction experiments were performed on the BT1 diffractometer at the NIST Center for Neutron Research (NCNR) using a Ge (311) monochromator at a wavelength of $\lambda = 2.0782$~\AA.  Rietveld refinements were carried out using the GSAS suite of programs.\cite{Rietveld}  The measurements were performed with two different cryostat/furnaces, one employed between $T = 10$~K and 580~K, and the other between $T = 575$~K and 675~K\@. \\ 

\begin{figure}[t]
\includegraphics[width=3.in]{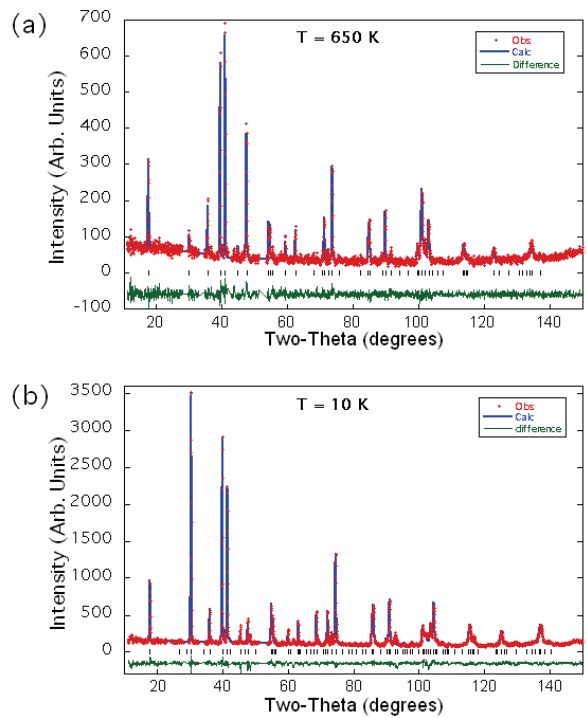}
\caption{(Color online) The observed neutron diffraction pattern (``+'' symbols) and fits from the Rietveld refinements (solid curves) at (a) $T$~=~650~K and (b) 10~K obtained for a powder sample of BaMn$_2$As$_2$.  The lower curves represent residuals from the refinement and the vertical bars are the expected peak positions.  More counts were accumulated at 10~K in order to obtain a better signal to noise ratio.}
\label{Fig-PND}
\end{figure}

Representative neutron diffraction patterns obtained at $T = 650$~K and $T = 10$~K are shown in Figs.~\ref{Fig-PND}(a) and (b) respectively, along with Rietveld refinements using the crystallographic (ThCr$_2$Si$_2$-type) and magnetic structures.  The magnetic structure was determined by a Rietveld refinement of the diffraction pattern at $T = 10$~K\@.  The magnetic structure was found to be $G$-type where the 4 Mn atoms situated at (1/2 0 1/4), (0 1/2 1/4), (0 1/2 3/4) and (1/2 0 3/4)  in the unit cell, are aligned antiferromagnetically in the sequence: up--down--up--down as shown in Fig.~\ref{Fig-structure}.  The results of the fit and the residuals of the refinement are shown in Fig.~\ref{Fig-PND}.  The refined value of the ordered moment at $T$~=~10~K is $\mu = 3.88(4)~\mu_{\rm B}$, which is somewhat lower than the value of $5.00~ \mu_{\rm B}$ expected for the high-spin ($S$~=~5/2) state of Mn$^{2+}$ assuming a $g$-factor $g = 2.00$ in the expression $\mu = gS\mu_{\rm B}$, where $\mu_{\rm B}$ is the Bohr magneton.

\begin{figure}[t]
\includegraphics[width=2.6in]{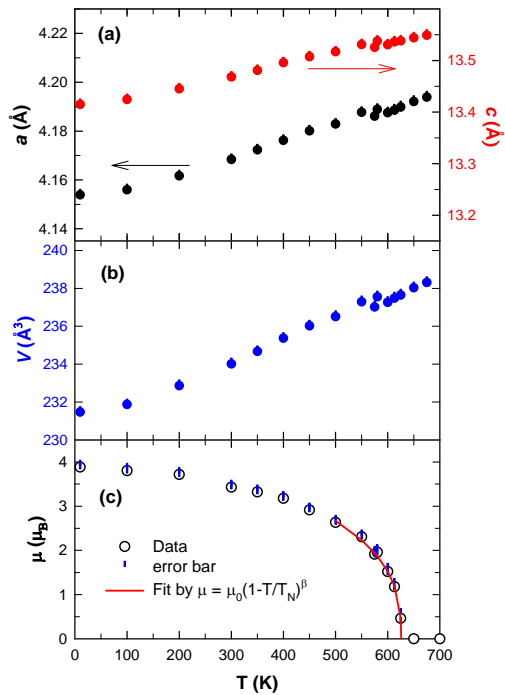}
\caption{(Color online) (a) Lattice parameters $a$ and $c$ versus temperature $T$, (b) unit cell volume $V$ versus $T$, and (c) ordered Mn moment $\mu$ versus $T$.  The solid curve through the data in (c) is a fit by the expression $\mu(T) = \mu_0(1-T/T_{\rm N})^{\beta}$.
\label{Fig-latticeparam(T)}}
\end{figure}

From Rietveld refinements of the diffraction patterns obtained at various temperatures using the ThCr$_2$Si$_2$-type crystal structure and the above $G$-type magnetic structure we obtained the variations of the lattice parameters $a$ and $c$, and the evolution of the ordered moment with temperature.  A selection of crystal and magnetic data fit parameters is shown in Table~\ref{atomicpositions}.  The variation of $a$, $c$, the unit cell volume $V = a^2c$, and the ordered moment $\mu$ versus $T$ are shown in Figs.~\ref{Fig-latticeparam(T)}(a), (b), and (c), respectively.  A monotonic increase in $a$ and $c$ with $T$ is observed, consistent with thermal expansion of the lattice.  The kinks at $T \approx 580$~K in the lattice parameters and unit cell volume versus $T$ most likely arise from a difference in temperature calibrations between the two sets of measurements.  The variation of $\mu(T)$ in Fig.~\ref{Fig-latticeparam(T)}(c) suggests that the antiferromagnetic transition is second order in nature.  Temperature scans were performed with increasing and decreasing temperature through $T_{\rm N}$ and no hysteresis was observed.  The data above $T = 500$~K were fitted by a power law $\mu(T) = \mu_0(1 - T/T_{\rm N})^{\beta}$ using $T_{\rm N}$, $\beta$ and $\mu_0$ as fitting parameters.  The fit shown as the solid curve through the data above $T = 500$~K in Fig.~\ref{Fig-latticeparam(T)}(c) gave the values $\beta = 0.35(2)$ and $T_{\rm N} = 625(1)$~K\@.  The crystal structure above and below $T_{\rm N}$ is the same tetragonal ThCr$_2$Si$_2$-type structure as demonstrated by the Rietveld refinements in Fig.~\ref{Fig-PND} and the fit parameters in Table~\ref{atomicpositions}.  

\begin{table*}

\caption{\label{atomicpositions}
Parameters obtained from Rietveld refinements of the powder neutron diffraction patterns at representative  temperatures.  Here $a$, $c$, and $V$ are the unit cell parameters and unit volume, respectively, $z$ is the As $z$ position in the crystal structure, $d_{\rm Mn-Mn}$ and $d_{\rm Mn-As}$ are the in-plane Mn--Mn and Mn--As distances, respectively, and $\mu$ is the refined ordered Mn moment.  The Rietveld refinements were performed in the \emph{I}4/\emph{mmm} space group with atomic positions Ba (0 0 0), Mn (1/2 0 1/4) and As (0 0 $z$).  A number in parentheses is the error (standard deviation) in the last digit of the respective quantity.}
\begin{ruledtabular}
\begin{tabular}{|c|c|c|c|c|c|c|c|c|c|}
Temperature & $a$ & $c$ & \emph{V} & \emph{z} &$d_{\rm Mn-Mn}$& $d_{\rm Mn-As}$& $\mu$& $R_{\rm wp}$ & $\chi^2$ \\
 (K) & (\AA) & (\AA) & (\AA$^3$)& & (\AA) & (\AA) & ($\mu_{\rm B}$/Mn) & (\%) & \\\hline  
10 & 4.1539(2) & 13.4149(8) & 231.47(2) & 0.3613(2) & 2.9373(2) & 2.558(2) & 3.88(4) & 7.81 &	1.69\\  
100& 4.1560(2) & 13.4246(8) & 231.87(2) & 0.3614(2) & 2.9387(2) & 2.560(2) & 3.80(5) & 8.87 & 1.12\\ 
200& 4.1617(2) & 13.4451(8) & 232.86(4) & 0.3612(2) & 2.9428(2) & 2.563(2) & 3.72(4) & 8.46 & 1.05\\ 
300& 4.1684(2) & 13.4681(8) & 234.01(4) & 0.3611(2) & 2.9475(2) & 2.566(2) & 3.43(4) & 8.31 & 0.964\\ 
400& 4.1762(2) & 13.4958(8) & 235.37(3) & 0.3611(2) & 2.9531(2) & 2.571(2) & 3.18(4) & 8.22 & 0.939\\ 
500& 4.1829(2) & 13.5171(8) & 236.51(4) & 0.3609(2) & 2.9578(2) & 2.573(2) & 2.63(4) & 8.73 & 0.951\\ 
600& 4.1875(2) & 13.5305(9) & 237.26(4) & 0.3618(3) & 2.9610(2) & 2.583(2) & 1.52(7) & 11.55 & 0.909\\ 
650& 4.1921(3) & 13.544(1) & 238.03(4) & 0.3616(3) & 2.9643(2) & 2.584(2) & NA & 13.15 & 0.942\\ 
\end{tabular}
\end{ruledtabular}
\end{table*}

The ordered moment $\mu = 3.88(4)~ \mu_{\rm B}$/Mn at $T = 10$~K suggests local moment antiferromagnetism in BaMn$_2$As$_2$.  This value is much larger than the values 0.2--1.0~$\mu_{\rm B}$/Fe found in the $A$Fe$_2$As$_2$ compounds.  However, this value is smaller than the value $\mu = 5~ \mu_{\rm B}$/Mn expected for the high-spin state of Mn$^{2+}$ moments as noted above.  A similar value $\mu = 4.2(1)~ \mu_{\rm B}$/Mn was observed for the isostructural phosphide BaMn$_2$P$_2$ which also has a high ordering temperature $T > 750$~K and an identical magnetic structure.\cite{Brock1994}  A reduced ordered moment is also observed in the compound La$_2$CuO$_4$.\cite{Johnston1997}  In the $A$Fe$_2$As$_2$ compounds, the reduced moment on Fe is attributed to the itinerant nature of the magnetism\cite{Yildirim2009} whereas in La$_2$CuO$_4$, the reduced ordered moment on Cu$^{+2}$ with spin $S = 1/2$ arises from strong quantum fluctuations due to the quasi-two-dimensionality of the square spin lattice.\cite{Johnston1997}  Recently, based on density functional calculations, An \emph{et al.}\ concluded that the $G$-type antiferromagnetic state is the most stable ground state for BaMn$_2$As$_2$ and calculated a reduced ordered moment $\mu = 3.35~ \mu_{\rm B}$/Mn.\cite{An2009}  From band structure calculations they find a substantial spin-dependent hybridization between the Mn 3$d$ and As 4$p$ orbitals and attribute the reduced ordered Mn moment to this hybridization.\cite{An2009}

A crucial difference between BaMn$_2$As$_2$ and the $A$Fe$_2$As$_2$ ($A$ = Ca, Sr, Ba, Eu) compounds is that a lattice distortion occurs at or near $T_{\rm N}$ of the $A$Fe$_2$As$_2$ compounds but none is observed at any temperature between 10~K and 675~K for BaMn$_2$As$_2$.  In the $A$Fe$_2$As$_2$ compounds, the AF structure was found to be a ``stripe'' structure within an approximately square spin lattice oriented along the $x$- and $y$-axes.  In the stripe structure, the magnetic structure of each of two sublattices defined by next-nearest-neighbor NNN Fe spins is a N\'eel (``G-type'') structure within the basal plane.  This results in nearest-neighbor NN spins being parallel along one basal-plane axis (the stripe direction, say $y$) and antiparallel along the other ($x$-direction).  Within a local moment Heisenberg interaction picture in the tetragonal structure, this magnetic structure has been shown theoretically to be stabilized if the fourfold AF NNN interactions $J_2$ are dominant over fourfold AF NN interactions $J_1$ such that $J_2 > J_{1}/2$.\cite{Yildirim2009}  In this case, the interaction between the two NNN sublattices is completely frustrated, such that the magnetic energy is independent of the easy-axis orientations of the two sublattices with respect to each other.  The total magnetic plus elastic energy of the system can be lowered if the frustration between the two NNN sublattices is reduced by a structural (orthorhombic) distortion within the $a$-$b$ plane which results in $J_{1x} \neq J_{1y}$.\cite{Yildirim2009}  This suggests that the observed tetragonal to orthorhombic structural distortions in the FeAs-based parent compounds are driven by magnetic interactions.  

To compare the exchange constants in (tetragonal) BaMn$_2$As$_2$ with those of the $A$Fe$_2$As$_2$ and LaFeAsO-type compounds, the comparison should be done with the FeAs-based compounds in their tetragonal phase (where they also do not show long-range magnetic order).  The tetragonal averages of the computed orthorhombic exchange constants for nine different FeAs-based parent compounds\cite{Yin2008} give $J_2 \gtrsim J_1/2$, consistent with the observed stripe order.  Inelastic neutron scattering studies of BaFe$_2$As$_2$ above $T_{\rm N}$ demonstrated that the wavevector of the antiferromagnetic fluctuations is the same as the ordering wavevector of the Fe spins in the stripe phase below $T_{\rm N}$,\cite{Matan2009} again indicating that $J_2 \gtrsim J_1/2$ in the tetragonal phase.  In BaMn$_2$As$_2$, on the other hand, the observed $G$-type magnetic structure can occur with NN interactions $J_1$ only, without any NNN $J_2$ interactions.  In particular, the distinctively different magnetic structures indicate that whereas NNN interactions are important in the tetragonal phase of Fe$_2$As$_2$-based compounds, NN interactions are dominant in BaMn$_2$As$_2$.

To summarize, we have found that in the ordered state the Mn spins ($S$~=~5/2) are aligned along the $c$ axis with an ordered moment $\mu = 3.88(4)~\mu_{\rm B}$ at $T = 10$~K\@.  The collinear magnetic structure is $G$-type where all NN spins are antiferromagnetically aligned.  This strongly contrasts with the stripe AF structure in the $A$Fe$_2$As$_2$ compounds, and indicates that within a local moment picture, NN exchange interactions are dominant in BaMn$_2$As$_2$ whereas NNN interactions are important in the corresponding tetragonal phase of the $A$Fe$_2$As$_2$ compounds.  The magnetic easy axis direction is along the $c$ axis, whereas it is in the $a$-$b$ plane in the $A$Fe$_2$As$_2$ compounds.  The magnetic transition was found to be second-order in nature as opposed to the first-order transition in one or more of the $A$Fe$_2$As$_2$ materials.  The difference between the experimentally observed ordering temperature from our neutron diffraction measurements [$T_{\rm N} = 625(1)$~K] and the value ($T_{\rm N} \sim 500$~K) previously extrapolated from magnetic susceptibility measurements\cite{Singh2009} evidently arises from inaccuracy in the earlier extrapolation.\cite{Singh2009}  Indeed, subsequent high-$T$ $\chi(T)$ measurements up to 950~K on similarly grown single crystals directly determined that $T_{\rm N} \approx 620$~K\@.\cite{Bella}  We infer that BaMn$_2$As$_2$, although isostructural to the tetragonal $A$Fe$_2$As$_2$ parent compounds, has magnetic and transport properties intermediate between those of these compounds and La$_2$CuO$_4$.  Suppression of the magnetic ordering in BaMn$_2$As$_2$ by doping might lead to materials with high superconducting transition temperatures.  The study of doped BaMn$_2$As$_2$ and further study of undoped BaMn$_2$As$_2$ is therefore of great interest as these studies might lead to new insights into the superconductivity in FeAs-based and La$_2$CuO$_4$-based compounds and related materials.

Work at the Ames Laboratory was supported by the Department of Energy - Basic Energy Sciences under Contract No.\ DE-AC02-07CH11358.  \\

\end{document}